\documentclass[a4paper,fleqn,usenatbib]{mnras}

\usepackage{savesym}
\usepackage{amsmath}
\savesymbol{iint}
\usepackage{txfonts}
\restoresymbol{TXF}{iint}

\usepackage[T1]{fontenc}
\usepackage{ae,aecompl}

\usepackage{graphicx}	% Including figure files
\usepackage{amsmath}	% Advanced maths commands
\usepackage{amssymb}	% Extra maths symbols
\usepackage{natbib}
\usepackage{subcaption,tikz}
\tikzset{boximg/.style={remember picture,red,thick,draw,inner sep=0pt,outer sep=0pt}}

\usepackage{float}
\definecolor{edits}{RGB}{255,127,0}
\usepackage{soul}
\sethlcolor{yellow}
\soulregister\citet7
\soulregister\citep7
\soulregister\citealp7
\soulregister\ref7

\renewcommand\hl[1]{#1}
\title[Magellanic origin for the Virgo substructure]{A Magellanic origin for the Virgo substructure}

\author[D. Boubert et al.]{
D. Boubert$^{1}$,\thanks{E-mail: d.boubert@ast.cam.ac.uk}
V. Belokurov$^{1,2}$,
D. Erkal$^{3}$, and
G. Iorio$^{1,4,5}$
\\
$^{1}$Institute of Astronomy, University of Cambridge, Madingley Road, Cambridge CB3 0HA, UK\\
$^{2}$Center for  Computational Astrophysics, Flatiron Institute, 162 5th Avenue, New York, NY 10010, USA\\
$^{3}$Department of Physics, Faculty of Engineering and Physical Sciences, University of Surrey, Guildford GU2 7XH, UK\\
$^{4}$Dipartimento di Fisica e Astronomia, Universit\`a di Bologna, via Gobetti 93/2, I-40129, Bologna, Italy\\
$^{5}$INAF - Osservatorio Astronomico di Bologna, via Gobetti 93/3, I-40129, Bologna, Italy
}

\date{Accepted XXX. Received YYY; in original form ZZZ}

\pubyear{2017}

\begin{document}
\label{firstpage}
\pagerange{\pageref{firstpage}--\pageref{lastpage}}
\maketitle

\begin{abstract}
\citet{iorio_first_2017} mapped out the Milky Way halo using a sample of RR Lyrae stars drawn from a cross-match of \emph{Gaia} with 2MASS. We investigate the significant residual in their model which we constrain to lie at Galactocentric radii $12<R<27\;\mathrm{kpc}$ and extend over $2600\;\mathrm{deg}^2$ of the sky. A counterpart of this structure exists in both the Catalina Real Time Survey and the sample of RR Lyrae variables identified in Pan-STARRS by \citet{2016ApJ...817...73H}, demonstrating that this structure is not caused by the spatial inhomogeneity of \emph{Gaia}. The structure is likely the Virgo Stellar Stream and/or Virgo Over-Density. We show the structure is aligned with the Magellanic Stream and suggest that it is either debris from a disrupted dwarf galaxy that was a member of the Vast Polar Structure or that it is SMC debris from a tidal interaction of the SMC and LMC $3\;\mathrm{Gyr}$ ago. If the latter then the sub-structure in Virgo may have a Magellanic origin.
\end{abstract}


%%%%%%%%%%%%%%%%%%%%%%%%%%%%%%%%%%%%%%%%%%%%%%%%%%

%%%%%%%%%%%%%%%%% BODY OF PAPER %%%%%%%%%%%%%%%%%%

\section{Introduction}
\label{sec:introduction}

It is the vast extent of the Galactic stellar halo that precludes a
comprehensive examination of all of its constituent
sub-structures. Stretched over tens, sometimes hundreds of kpc, faint
stellar streams enter our view only briefly to be cut off abruptly by
the limitations of a survey, even the most ambitious one. In fact,
until the present day no single-instrument optical experiment has been
truly all-sky. With \emph{Gaia} Data Release 1 out more than a year ago
\citep[see][]{gaia_collaboration_gaia_2016-1,gaia_collaboration_gaia_2016,gaia_dr1_photometry_2017} -
and the DR2 recently released - it is a whole
new ball game.  \emph{Gaia} appears to have been built as a nearly
perfect halo exploration machine: what it lacks in depth (as it
``only" reaches r $\sim$ 21) it more than makes up in coverage (entire
sky), exquisite resolution (on par with the HST), quality of
photometry (given the enormous CCD array a star has to cross),
spectral coverage (most of the optical wavelength range), variability
data (with $\sim$70 visits per location during the mission lifetime)
and artefact rejection. With its uniform whole sky coverage \textit{Gaia} will enable more confident identification of diffuse halo structures than was possible previously, while its astrometry will enable the determination of their origin.

In the halo, where 99\% of the available Galactic volume is filled
with a measly 1\% of the available light, the surface brightness levels
in question are prohibitively faint \citep[see e.g.][]{BJ_2005} and
thus the choice of the tracer population is crucial \citep[see
  e.g.][]{Belokurov_dwarf_review_2013}. RR Lyrae stars (RRLs) - pulsating helium
burners - have been the weapon of choice for stellar halo studies for
decades \citep[see
  e.g.][]{hawkins1984,saha1984,wm1996,ivezic2000,Vivas,watkins,S82,rrl3,torrealba_discovery_2015,oglerrl2016,belokurov_clouds_2017}. Located
on the horizontal branch \citep[see e.g.][]{catelan2009}, these old
and metal-poor stars occupy a narrow range of intrinsic luminosities
and suffer minuscule amounts of contamination (if variability
information is on hand). %RRLs exhibit a relationship between the
%pulsation period and luminosity. However, the slope of this
%relationship changes quickly as a function of wavelength and
%stabilises only in the IR \citep[see e.g.][]{Madore_2013}. In the
%white light - i.e. V band or \emph{Gaia}'s G - the period-luminosity
%relationship is actually rather flat \citep[see][]{Catelan_2004}, thus
%obfuscating the need for period determination.

\begin{figure*}
	\includegraphics[width=\textwidth,trim = 1mm 27mm 15mm 30mm, clip]{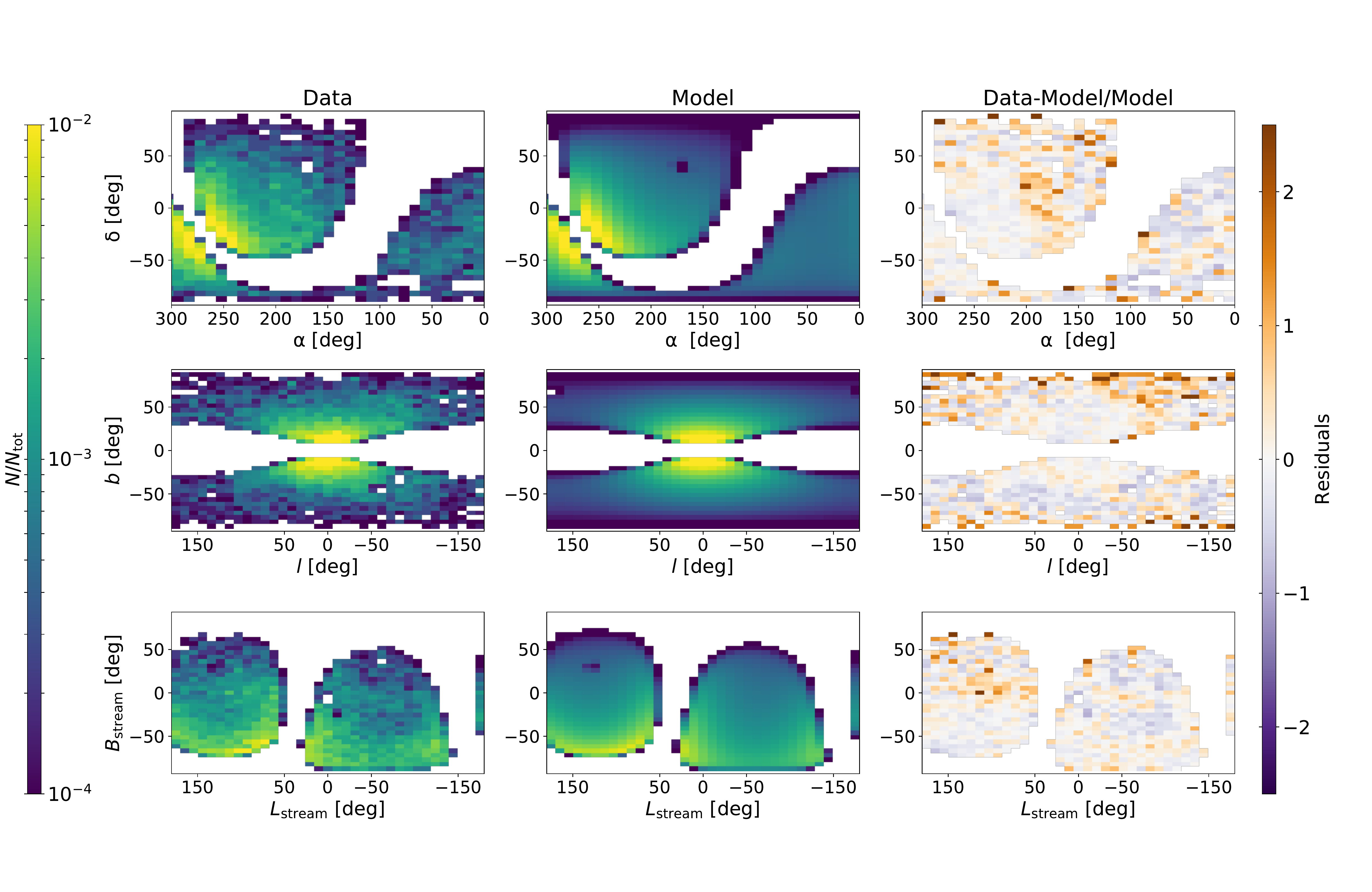}
	\caption{Comparison of the \emph{Gaia}+2MASS RRLs dataset to the best-performing halo model of \citet{iorio_first_2017} and the residuals. There is one prominent residual that spans from $(\alpha,\delta)=(160^{\circ},-50^{\circ})$ to $(200^{\circ},30^{\circ})$ in equatorial coordinates, from $(l,b)=(-80^{\circ},20^{\circ})$ to $(-60^{\circ},70^{\circ})$ in Galactic coordinates, and from $(L_{\mathrm{MS}},B_{\mathrm{MS}})=(50^{\circ},0^{\circ})$ to $(120^{\circ},0^{\circ})$ in Magellanic Stream coordinates. \textbf{Top:} Equatorial coordinates. \textbf{Middle:} Galactic coordinates. \textbf{Bottom:} Magellanic stream coordinates but with the pole moved to $(l,b)=(187^{\circ},8^{\circ})$ (see Sec. \ref{sec:crts}). This figure is from \citet{iorio_first_2017} but with additional panels to highlight the substructure in the adapted Magellanic Stream coordinates. }
	\label{fig:modelresiduals}
 \end{figure*}

In a recent work \citet{iorio_first_2017} combined \emph{Gaia} DR1 and Two Micron All Sky Survey (2MASS) photometry to obtain a collection of $\approx 21,600$ RRLs out to $\approx 20\;\mathrm{kpc}$ and used that sample to fit a model of the Milky Way halo.  Taking advantage of the stable completeness and the all-sky view of their sample, \citet{iorio_first_2017} were able to test a suite of sophisticated models of the Milky Way stellar halo. While these models captured the variation of the triaxiality with Galactocentric radius as mapped out by the RRLs, a significant residual remained in the region $\{l\in(-100^{\circ},-50^{\circ}), b>0^{\circ}, d>10\;\mathrm{kpc}\}$ (see Fig. \ref{fig:modelresiduals}). \citet{iorio_first_2017} proposed that this residual was the substructure identified as the Virgo Over-Density (VOD).

The VOD is a diffuse cloud of stars covering at least $2000\;\mathrm{deg}^2$ of the sky \citep{2012AJ....143..105B} and with an apparent extent along the line of sight of $5\;\mathrm{to}\;20\;\mathrm{kpc}$ \citep{juric_milky_2008}. The true extent of the VOD is unknown due to the richness of substructure near Virgo; one wrap of the Sagittarius stream is present at between $30\;\mathrm{and}\;60\;\mathrm{kpc}$ \citep{vivas_disentangling_2016} and the more distant portion of the VOD may be a distinct piece of substructure known as the Virgo Stellar Stream (VSS, e.g. \citealp{2006ApJ...636L..97D}). \citet{2012AJ....143..105B} found that the VOD peaks at. $d_{\odot}\sim 11 \;\mathrm{kpc}$ and is detected as nearby as $d_{\odot}=7\;\mathrm{kpc}$, while \citet{zinn_silla_2014} found the VSS is centered at $d_{\odot}\sim 19\;\mathrm{kpc}$ but extends across the range $17<d_{\odot}<22\;\mathrm{kpc}$. It is likely that the residual in the model of \citet{iorio_first_2017} is the VOD and/or the VSS, but we present here a new view of the Virgo sub-structure as a stream leading the Magellanic Clouds.

In Section \ref{sec:samples} we isolate the substructure in the
\emph{Gaia}--2MASS sample of \citet{iorio_first_2017} as well as in the RRL sample of the Catalina Real-Time Transient Survey (CRTS) and in Pan-STARRS \citep{2016ApJ...817...73H}. We discuss the connection of the structure to the VOD and VSS in Section \ref{sec:discussion}, as well as presenting a Magellanic interpretation of its origin.

\section{RR Lyrae star samples}
\label{sec:samples}

\subsection{Gaia+2MASS}
\label{sec:gaia2mass}

The selection of RRLs by \citet{iorio_first_2017} relied on the color provided by the \emph{Gaia} and 2MASS cross-match $J-G$ and the proxy of stellar variability, $\mathrm{AMP}$,  proposed by \citet{belokurov_clouds_2017} and \citet{deason_clouds_2017},
\begin{equation}
\mathrm{AMP}\equiv \log_{10}\left(\sqrt{N_{\mathrm{obs}}}\frac{\sigma_{\mathrm{F}_G}}{\mathrm{F}_G}\right),
\end{equation}
where $N_{\mathrm{obs}}$ is the number of times a source has crossed a CCD in Gaia's focal plane, $\mathrm{F}_G$ is the flux (electron per second) measured in the $G$ band averaging over $N_{\mathrm{obs}}$ single flux measurements, and $\sigma_{\mathrm{F}_G}$ is the standard deviation of the $N_{\mathrm{obs}}$ flux measurements. \citet{iorio_first_2017} demonstrated that suitable cuts in both $\mathrm{AMP}$ and $J-G$ colour were able to select a sample where the contamination is expected to be substantially less than 10\%. Importantly, they also argued that the completeness of their selection did not vary strongly, either as a function of the position on the sky or the magnitude of the star.

We use the cuts given in Table 1 of \citet{iorio_first_2017}. For this work the most relevant cuts are the magnitude cut ($10<G<17.1$, \hl{corresponding to heliocentric distances $10<d_{\odot}<20\;\mathrm{kpc}$}) and the structure cuts which were made to exclude both Magellanic Clouds in addition to the regions S1 $\{l\in(167^{\circ},189^{\circ}), b\in(16^{\circ},22^{\circ})\}$ and S2 $\{l\in(160^{\circ},190^{\circ}), b\in(63^{\circ},73^{\circ})\}$, which \citet{iorio_first_2017} found to contain spurious extended structure speculated to be due to \emph{Gaia} cross-match errors.

\citet{iorio_first_2017} found that a density profile with a single power law in radius, flattened along the $Z$-axis with the flattening varying with radius, was able to well describe the density distribution of RRLs. In Figure \ref{fig:modelresiduals} we compare this model to the distribution of \emph{Gaia}+2MASS RRLs. There is a remarkable spatially-localised residual spanning $70^{\circ}$ whose coordinates are given in the caption of Figure \ref{fig:modelresiduals}. While \citet{iorio_first_2017} did exclude the regions S1 and S2 from their sample because of suspected cross-match errors with the \emph{Gaia} source catalogue, this structure is unlikely to be due to similar errors because of its large spatial extent. \citet{iorio_first_2017} suggested the structure may be related to the Virgo Over-Density and we discuss this further in Section \ref{sec:discussion}.

Intriguingly, the structure is almost aligned with the leading portion of the Magellanic Stream. To test this we define a coordinate system $(L_{\mathrm{stream}},B_{\mathrm{stream}})$ where the equator is aligned with the structure and the LMC lies at $L_{\mathrm{stream}}=0$, giving a pole in Galactic coordinates of roughly $(l,b)=(187^{\circ},8^{\circ})$. This coordinate system is thus almost aligned with the Magellanic Stream coordinates defined by \citet{nidever_origin_2008} who found a pole of $(l,b)=(188.5^{\circ},7.5^{\circ})$ by fitting a great circle to the gas in the Magellanic Stream. The Magellanic Stream has a width of $20^{\circ}$ (see Fig. 8 in \citealp{nidever_origin_2008}) and thus this difference is negligible.

Note that the selection cut in magnitude $G<17.1\;\mathrm{mag}$, used by \citet{iorio_first_2017} to ensure that the completeness did not vary with distance, corresponds to a maximum distance of $20.7\;\mathrm{kpc}$ (assuming an absolute magnitude $M_G=0.52$ for the RRL stars); to investigate whether the structure continues to greater distances we must turn to either \textit{Gaia} DR2 or one of the ground-based RRL surveys.

\subsection{CRTS}
\label{sec:crts}

The Catalina Sky Survey (CSS) is a synoptic survey that has covered more than $33,000\;\deg^2$ of the sky utilising three telescopes. The Catalina Real-Time Transient Survey (CRTS) analyses the data from the CSS to identify optical transients. The CRTS has catalogued RRLs in both the Northern hemisphere \citep{drake_evidence_2013,drake_probing_2013,drake_catalina_2014} and Southern hemisphere \citep{torrealba_discovery_2015,2017MNRAS.469.3688D}. Bringing together the various published catalogues we have a sample of 31301 RRLs, after removing duplicates. These stars cover almost the entire sky, excluding the region $|b|\lesssim 20^{\circ}$ near the Galactic disc, and extend out to $d\lesssim 70\;\mathrm{kpc}$.

\begin{figure*}
	\includegraphics[width=1.0\textwidth,trim = 26mm 2mm 35mm 13mm, clip]{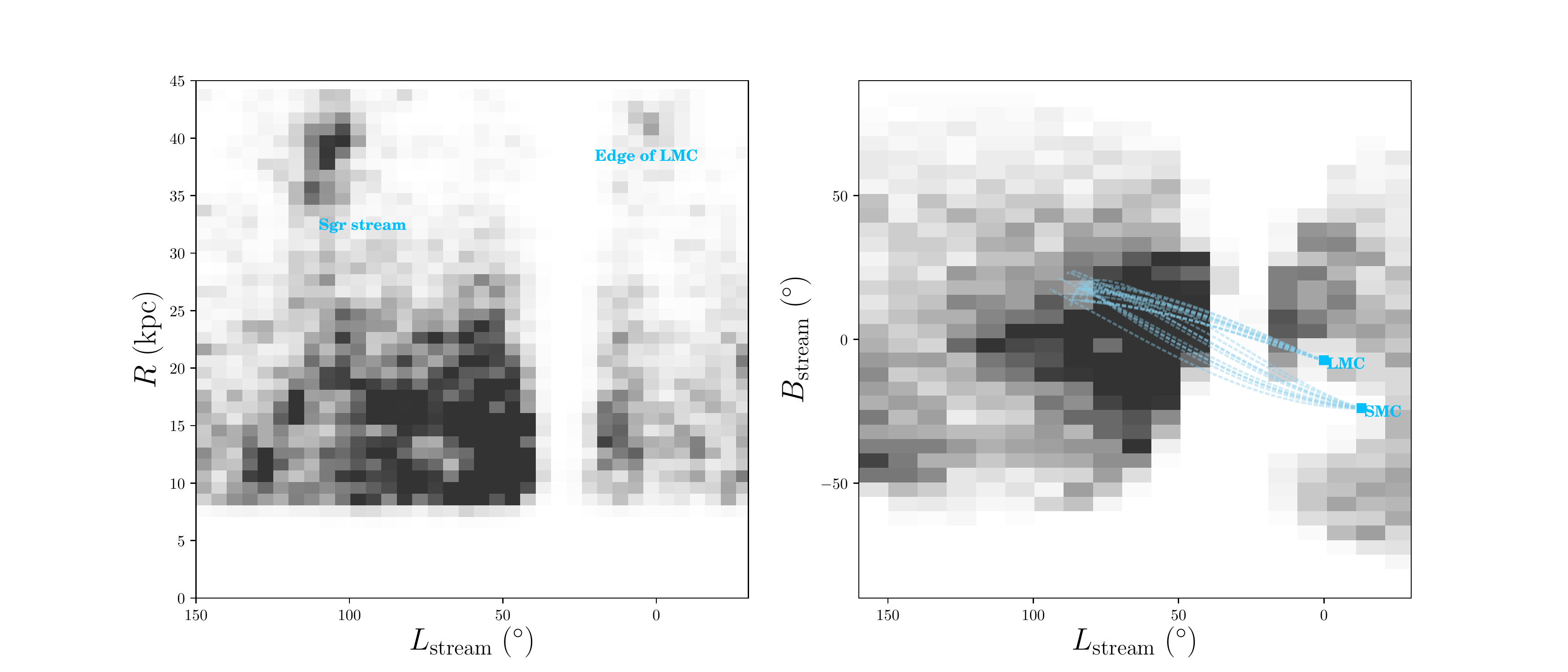}
	\caption{\textbf{Left:} Distribution of CRTS RR Lyrae stars (RRLs) in distance versus modified Magellanic Stream longitude with $|B_{\mathrm{stream}}|<8^{\circ}$. \textbf{Right:} Sky distribution of the subset of CRTS RRLs in the Galactocentric radius range $12<R<27\;\mathrm{kpc}$. The ten dashed lines are orbital tracks for the LMC and SMC from present day to $1\;\mathrm{Gyr}$ in the future (described further in Sec. \ref{sec:magstream}).}
	\label{fig:crtsstream}
 \end{figure*}

In the left panel of Fig. \ref{fig:crtsstream} we show the distribution in $L_{\mathrm{stream}}$ of the CRTS RRLs that lie within $-8^{\circ}<B_{\mathrm{stream}}<8^{\circ}$. Both the Sagittarius stream and the edge of the LMC are clearly visible beyond $35\;\mathrm{kpc}$, but closer in there is an extended structure ranging over $40^{\circ}<L_{\mathrm{stream}}<110^{\circ}$ and $8<R<27\;\mathrm{kpc}$. There is evidence of a gap at $12\;\mathrm{kpc}$ and we discuss whether this is the divide between the VOD and the VSS in Sec. \ref{sec:discussion}. The rapid drop-off in contamination from the smooth stellar halo makes it easier to study the more distant structure and thus we focus on the subset of RRLs with Galactocentric radii $12<R<27\;\mathrm{kpc}$.

 \begin{figure}
 	\centering
 	\begin{subfigure}{0.5\textwidth}
 		%\centering
 		\includegraphics[width=1.0\textwidth,trim = 0mm 0mm 10mm 10mm, clip]{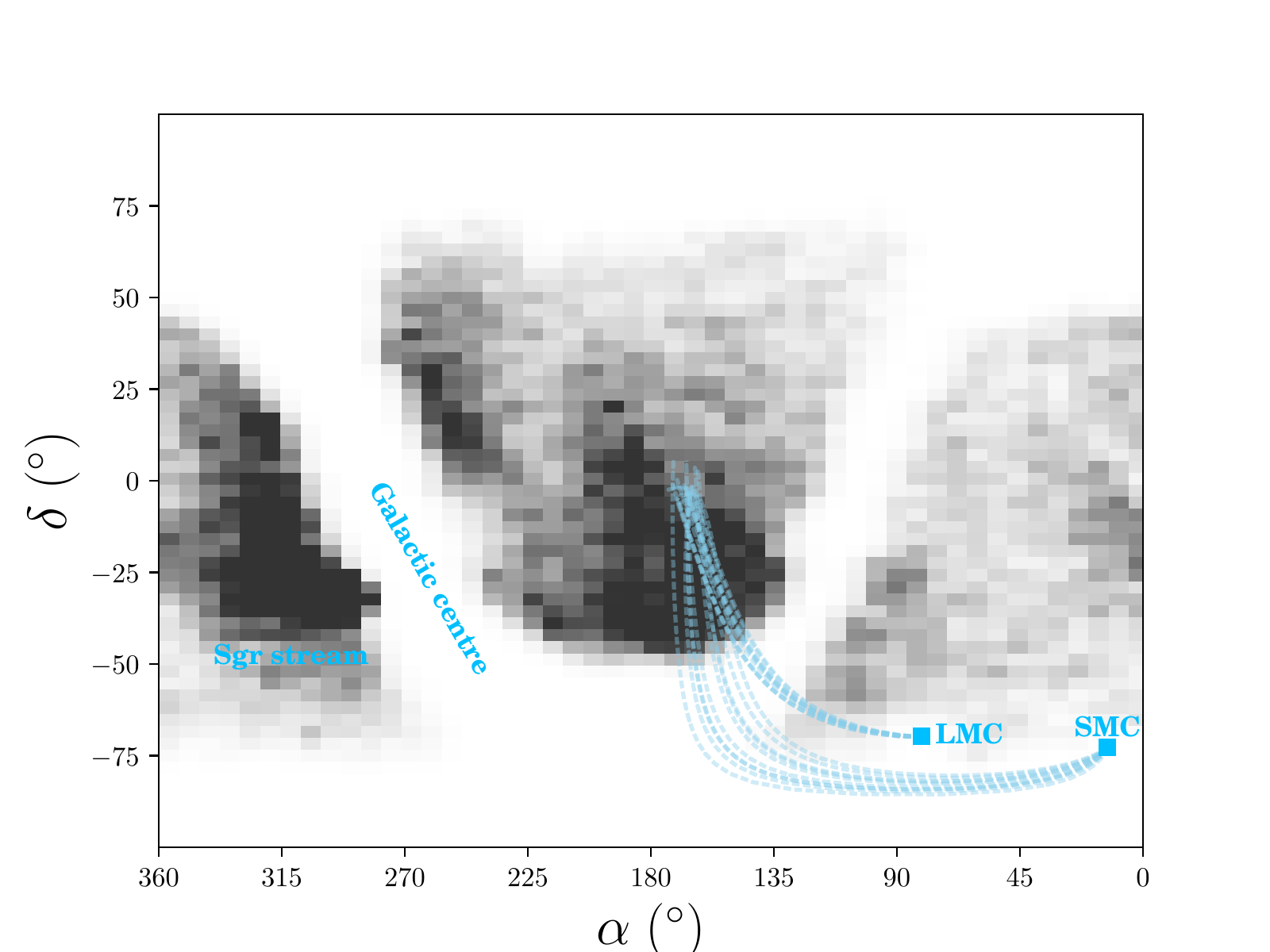}
 		\caption{CRTS}
 		\label{fig:crtsradec}
 	\end{subfigure}
 	\begin{subfigure}{0.5\textwidth}
 		%\centering
 		\includegraphics[width=1.0\textwidth,trim = 0mm 0mm 10mm 10mm, clip]{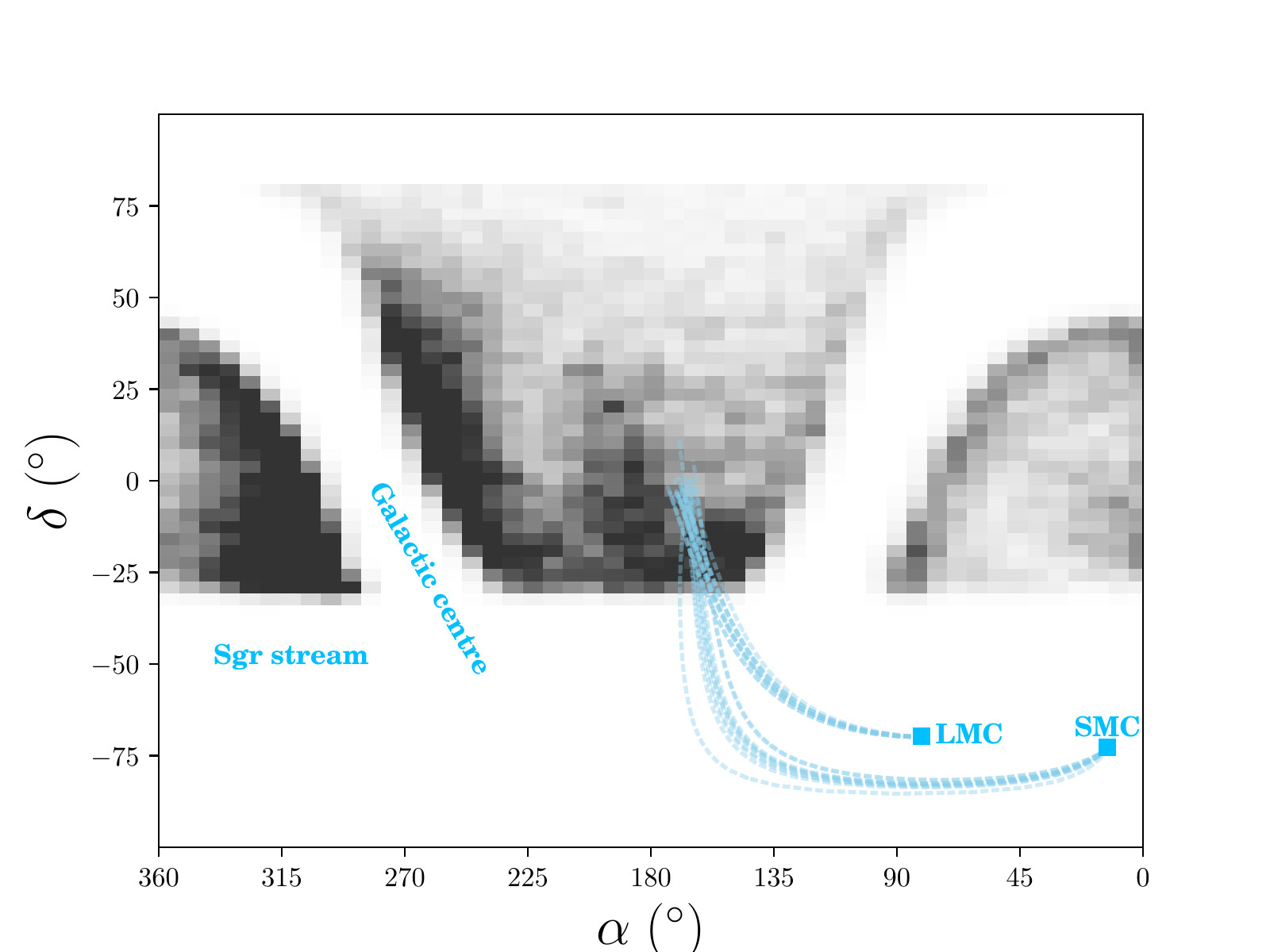}
 		\caption{Pan-STARRS}
 		\label{fig:panstarrsradec}
 	\end{subfigure}
	
	\caption{The distribution of the RRLs in both the CRTS and Pan-STARRS lying in the Galactocentric radius range $12<R<27\;\mathrm{kpc}$ plotted in equatorial coordinates. The locations of the LMC and SMC are shown in addition to ten randomly sampled orbit tracks from present day to $1\;\mathrm{Gyr}$ in the future. The other regions of high density correspond to the Galactic centre and a nearby portion of the Sagittarius stream. The orbital tracks are as described in Fig. \ref{fig:crtsstream}.}
	\label{fig:bothradec}
\end{figure}

There are 14752 RRLs in the CRTS with Galactocentric radius, $R$, lying in the range $12<R<27\;\mathrm{kpc}$ and we plot their distribution on the sky in Fig. \ref{fig:crtsradec}. The structure found in the \emph{Gaia}+2MASS sample has a clear counterpart in the CRTS and thus cannot be an artefact due to internal \emph{Gaia} cross-match errors. \hl{We show that the CRTS RRLs are aligned with the stream coordinate system chosen for the \textit{Gaia}-2MASS RRLs in the right panel of Fig. \ref{fig:crtsstream}, which confirms that the structures found in each sample are the same.}

Ideally, we would have accurate metallicities for the CRTS RRLs because this would give further information on their origin. Indeed, it is possible to estimate the metallicity of an RRL using the light curve \citep{1995A&A...293L..57K,1996A&A...312..111J} and \hl{we estimate such metallicites for the CRTS RRLs using the procedure described in Sec. 4.2 of \citet{torrealba_discovery_2015}}. However, cross-matching with \hl{SDSS DR12} for the Northern tip of the stream reveals that the the light curve metallicities are offset from the SDSS metallicities by $-0.29\pm0.55\;\mathrm{dex}$ and so are not reliable estimators of the true metallicities. Inspection of the radial velocities and metallicities of the 1029 successfully cross-matched stars did not reveal any obvious sub-population of unusually fast-moving or metal-poor stars, which we tentatively conclude implies that few stars in this subset are members of the structure. \hl{A major caveat to this conclusion is that the radial velocity of RRLs can be shifted by up to $100\;\mathrm{km}\;\mathrm{s}^{-1}$ by the pulsations (see \citealp{vivas_disentangling_2016} for a proper treatment) and thus the velocity signal of the structure could be blurred out.}

\subsection{Pan-STARRS}
\label{sec:panstarrs}

The Pan-STARRS Collaboration is using a synoptic imaging system to produce surveys covering all distance scales from Solar System to extragalactic \citep{2016arXiv161205560C}. The deep, large-area component is the $3\pi$ Steradian Survey which has obtained broadband $grizy$ photometry over the region $\delta>-30^{\circ}$ and down to $g\sim23$ in the stacked images. The completeness of the $3\pi$ survey is known to be spatially heterogeneous in the first data release \citep{2016arXiv161205560C}, however this heterogeneity should be independent of the heterogeneities of our \textit{Gaia}+2MASS and CRTS samples. \citet{2016ApJ...817...73H} used the 35 epochs from the first data release to assign a probability $P_{\mathrm{RRL}}$ to each source of it being an RRL based on the amplitude and timescales of variability in the lightcurves. \citet{2016ApJ...817...73H} found that by applying the cut $P_{\mathrm{RRL}}>0.2$ they achieved a purity of $\sim75\%$ and a completeness of $~92\%$ relative to SDSS Stripe 82. In this work we use the 42,674 ``highly likely halo RR Lyrae candidates" \citep{2016ApJ...817...73H} defined by $|b|>20^{\circ}$ and $P_{\mathrm{RRL}}>0.2$. Applying a Galactocentric radius cut, $12<R<27\;\mathrm{kpc}$, reveals the same structure that is present in the \emph{Gaia}+2MASS and CRTS samples (see Fig. \ref{fig:panstarrsradec}).

\section{Discussion}
\label{sec:discussion}

Identification of the structure in each of \emph{Gaia}+2MASS, CRTS and Pan-STARRS makes it extremely likely that the structure is real. The structure lies in a region of the sky and at a distance that has previously been identified with the Virgo Over-Density (VOD) and/or Virgo Stellar Stream (VSS) and we discuss their relationship in Section \ref{sec:possibilities}. The near alignment with the Magellanic Stream is suggestive that the structure may have a Magellanic origin and we expand on this possibility in Section \ref{sec:magstream}. However, we are unable to conclusively determine the structure's origin because we only have positions and distances. 

\subsection{Comparison of the three datasets}
\label{sec:comparison}
The three RRL datasets are complementary in their differing balance between contamination and completeness. The CRTS has close to zero contamination, but the completeness is a function of magnitude and position on the sky due to the varying number of observations per field \citep{drake_probing_2013}. On the other hand, the RRLs in the 2MASS+Gaia sample has a high rate of contamination, but the completeness was shown by \citet{iorio_first_2017} to not vary significantly either spatially or with magnitude up to $G<17.1\;\mathrm{mag}$. Somewhere between these two samples lies the \citet{2016ApJ...817...73H} sample of RRLs in the first data release of Pan-STARRS, which is known to be spatially heterogeneous \citep{2016arXiv161205560C} in a way that is independent of either of the other two samples. \citet{2016ApJ...817...73H} showed that their sample is $\sim75\%$ pure and $\sim92\%$ complete relative to SDSS Stripe 82 RRLs.

\subsection{Characterization of the structure}
\label{sec:structure}
To aid the following discussion we briefly characterize the structure.

\begin{description}
	\item[\textbf{Outer edge:}] We can not say if we have identified the outer edge of the structure. The left panel of Fig. \ref{fig:crtsstream} shows a rapid decrease in RRL number density past $27\;\mathrm{kpc}$, however we would expect the completeness of CRTS to decrease with increasing distance. The Sgr stream is clearly defined beyond $35\;\mathrm{kpc}$ in both our CRTS and Pan-STARRS samples, however this may merely be due to the large numbers of stars expected in that structure; the Sgr dwarf was originally similar in mass to the SMC (e.g. \citealp{2010ApJ...712..516N}) and thus even if we only see 1\% of RRLs at $50\;\mathrm{kpc}$ that still equates to a large number of stars. It is possible that the more distant portions of this structure, which we should expect to contain many times fewer stars than the Sgr dwarf, may simply be drowned out.
    \item[\textbf{Distance gradient:}] If the structure had a gradient in distance along its length, then we may be able to use that gradient to trace the origin of the structure despite not having velocities. In Fig. \ref{fig:crtsstream} there does appear to be a trend in the most prominent part of the structure of decreasing Galactocentric radius as the structure approaches the plane, however if the more distant parts of the structure are included then this gradient disappears.
    \item[\textbf{Area:}] To estimate the area of the structure, we drew a contour at a stellar density corresponding to $0.5\;\mathrm{stars}\;\deg^{-2}$ (chosen by eye to select the structure) and calculated the area contained in the contour. The resulting area of roughly $2600\;\deg^{-2}$ corresponds to one-sixteenth of the sky or $\frac{\pi}{4}$ steradians. We find that 2290 of the RRLs stars with $12<R<27\;\mathrm{kpc}$ are inside this contour, however, we caution that this is only an estimate of the number of the stars in the structure due to both the uncertainty in the shape of the contour and contamination from the smooth stellar halo.
    \item[\textbf{Gap:}] \hl{The left-hand panel of Figure 2 shows a gap in the structure between $60<L_{\mathrm{stream}}<80^{\circ}$. We select the stars with $R<20\;\mathrm{kpc}$ in this region (in addition to the cut $|B_{\mathrm{stream}}|<8^{\circ}$) and calculate a kernel density estimate as a function of the Galactocentric radius. The density shows two clear peaks of comparable height at $10\;\mathrm{kpc}$ and $16\;\mathrm{kpc}$, with a trough at $12.5\;\mathrm{kpc}$ that corresponds to the gap. The minima of the trough is 30\% below the maxima of the peaks and thus the gap is a significant deviation from a singly-peaked structure.}
\end{description}

\subsection{Relationship to the VOD/VSS}
\label{sec:possibilities}

\citet{iorio_first_2017} suggested that the residual in their model was related to the substructure in Virgo, specifically the VOD, and this is confirmed by the distance distribution shown in Fig. \ref{fig:crtsstream}. There is an extensive literature surrounding the sub-structure in Virgo which is well-reviewed by \citet{2016ASSL..420...87G}. First revealed by \citet{2002ApJ...569..245N} in a single stripe of SDSS, the large extent of the VOD has long been established both on the sky (more than $1000\;\deg^2$, \citealp{juric_milky_2008}, more than $2000\;\deg^2$, \citealp{2012AJ....143..105B}) and along the line-of-sight (a typical distance range of $10<d_{\odot}<20\;\mathrm{kpc}$ was obtained using RRLs from both the QUEST survey, \citealp{2001ApJ...554L..33V,2003MmSAI..74..928V,Vivas}, and the SEKBO survey, \citealp{2008ApJ...678..851K,2009MNRAS.394.1045K}).

The VOD has already been studied using the CRTS RRL; \citet{torrealba_discovery_2015} applied the \citet{1980ApJ...236..351D} clustering algorithm to the CRTS RRL sample and identified 12 structures that had a significance greater than $3\sigma$ (shown in their Fig. 14). \citet{torrealba_discovery_2015} noted that their second most significant and largest by area cluster, Hya 1, may be a blend between structure around the Galactic centre and stars in the halo. The part of Hya 1 away from the plane and the other over-densities Cen 1, Cen 5, Cen 6 and Cen 7 are consistent with being part of the nearer structure which may be the VOD. The over-densities Hya 4, Cen 2, Cen 3 and Cen 4 are consistent with the structure we have referred to as the VSS. \citet{torrealba_discovery_2015} do suggest that some of these over-densities could be related to the VOD, but this proposal is not discussed in detail.

More recently, \citet{vivas_disentangling_2016} assembled a sample of 412 RRLs with radial velocities and distances $4<d_{\odot}<75\;\mathrm{kpc}$ which cover the region $175^{\circ}<\alpha<210^{\circ}$, $-10^{\circ}<\delta<+10^{\circ}$. \citet{vivas_disentangling_2016} identified six significant groups, including the VSS with 18 stars across $15.6<d_{\odot}<21.3\;\mathrm{kpc}$ ($16.5<R_{\mathrm{GC}}<21.8\;\mathrm{kpc}$) and a mean heliocentric radial velocity of $135.2\;\mathrm{km}\;\mathrm{s}^{-1}$. The region considered by \citet{vivas_disentangling_2016} lies at the very tip of the structure and their VSS group occupies the same range of Galactic radii.

\hl{The VOD and VSS may be two components of a single structure (for instance \citealp{zinn_silla_2014} showed that the nearest portions of the VSS overlap with the VOD). A likely interpretation of the gap shown in the left-hand panel of Fig. \ref{fig:crtsstream} and quantified in Sec. \ref{sec:structure} is that it is the gap between the VOD and VSS. The gap appears to end at $L_{\mathrm{stream}}=60^{\circ}$, which leaves open the possibility of the VOD and VSS merging into a single structure at low longitudes (low Galactic latitudes). Further work will be required using the proper-motions in \textit{Gaia} DR2 to check whether this join is merely overlap (as cautioned by \citealp{zinn_silla_2014}).}

In summary, the residual mapped out in Sec. \ref{sec:samples} is almost certainly associated with the VOD/VSS. If our structure is part of the VSS/VOD then these stars will have likely originated in a dwarf galaxy that has been disrupted by the Milky Way \citep[e.g.][]{2012AJ....143..105B,2016ASSL..420...87G}. The alignment of this structure with the orbital plane of the Magellanic Clouds would then be naturally explained by the progenitor being a member of the well-known Vast Polar Structure of satellite galaxies of the Milky Way \citep{2012MNRAS.423.1109P}. Conversely the structure may be two distinct substructures which only appear adjacent on the sky, with the VSS at the tip and the lower latitude portion having an alternative origin. 

\subsection{Near alignment with the Magellanic stream}
\label{sec:magstream}

The Large and Small Magellanic Clouds are the two most massive dwarf galaxies in the vicinity of the Milky Way and have a history of interaction that has likely lasted for several $\mathrm{Gyr}$. Evidence for these interactions includes the $200^{\circ}$ gaseous Magellanic Stream and Leading Arm \citep{2010ApJ...723.1618N} and both the gaseous \citep{1963AuJPh..16..570H} and stellar \citep[i.e.][]{belokurov_clouds_2017} components of the Magellanic Bridge which stretches between the two galaxies. During these interactions, stars will be stripped from the outskirts of the SMC and form a stellar counterpart to the Magellanic Stream.

\begin{figure}
	\includegraphics[width=0.5\textwidth,trim = 0mm 0mm 10mm 10mm, clip]{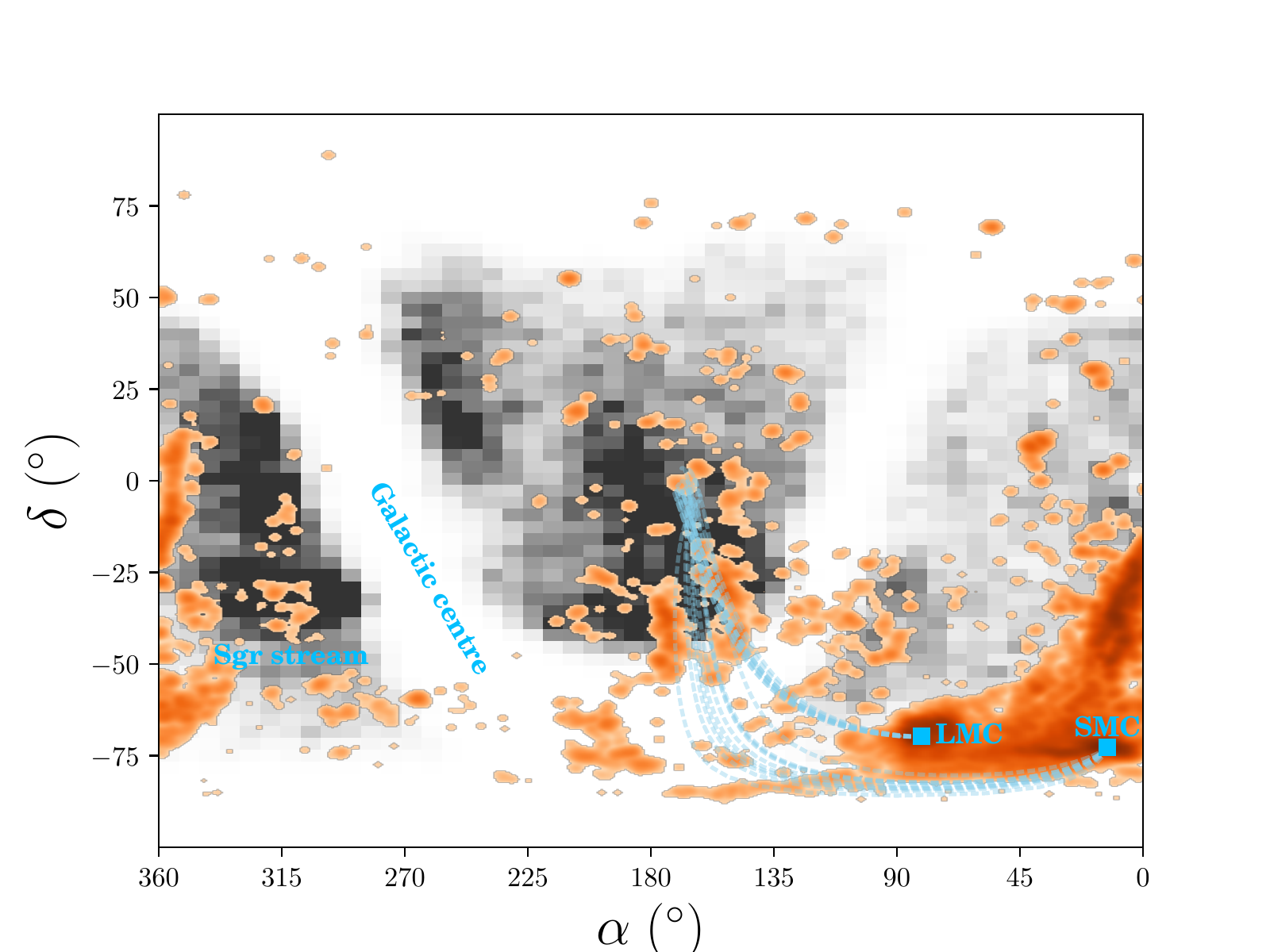}
	\caption{The underlying density and labels are the same as Fig. \ref{fig:crtsradec}. Overplotted is the GASS HI map of the Magellanic Stream from \citet{2010ApJ...723.1618N}.}
	\label{fig:magstream}
 \end{figure}

The structure we have identified is nearly aligned with the Magellanic Stream and thus we must consider the possibility of a Magellanic origin. \citet{2010ApJ...723.1618N} compiled a map of the $200^{\circ}$ long Magellanic Stream in H{\sc I}; in Figure \ref{fig:magstream} we demonstrate that the leading arm in this map is coincident with the CRTS RRLs with Galactocentric radii $12<R<27\;\mathrm{kpc}$. One explanation for the Magellanic Stream is that it is composed of gas stripped from the SMC by the tidal action of the LMC during previous close interactions and that same tidal action should also strip stars from the SMC. These stars may then have similar kinematics to the stripped gas. \citet{2008ApJ...673L.143M} used the interaction of a high-velocity cloud in the Leading Arm with the Galactic disc to estimate that the Leading Arm crosses the Galactic plane at a Galactic radius of $R_{\mathrm{GC}}\approx 17\;\mathrm{kpc}$. \citet{2012MNRAS.421.2109B} modeled the evolution of the Magellanic Stream and, in their Model 2 with a significant Leading Arm, they found that the distance gradient of the material was flat over Magellanic Stream latitudes $20<L_{\mathrm{MS}}<80\;\deg$. Thus the Galactic plane distance $R_{\mathrm{GC}}\approx 17\;\mathrm{kpc}$ calculated by \citet{2008ApJ...673L.143M} may be directly compared to the distance distribution $12<R_{\mathrm{GC}}<27\;\mathrm{kpc}$ used to isolate the RRL structure, even though we are examining the structure at high latitude $b>20^{\circ}$.

%An implication of the Magellanic Stream having both a gaseous and stellar component is that it should host a substantial dark matter component. The mass of the (atomic+ionized) gaseous Magellanic Stream has been estimated at around $2\times 10^9\;\mathrm{M}_{\odot}$ \citep{2014ApJ...787..147F} and so assuming the cosmological baryon fraction $15.8\%$ \citep{2016A&A...594A...1P} we would infer $10^10\;\mathrm{M}_{\odot}$ of dark matter which is greater than the mass of the SMC. As much as $10^9\;\mathrm{M}_{\odot}$ of this may be in the Leading Arm, which if it crosses the Galactic plane at $R_{\mathrm{GC}}$ as discussed above would leave a significant imprint in the kinematics of the Milky Way near the crossing point.

There are other possible explanations involving the Magellanic Clouds. The stars may have been tidally stripped from the LMC by the Milky Way if the Magellanic Clouds have experienced more than one perigalactic passage. There are several mechanisms that could eject stars from the Magellanic Clouds (see \citealp{boubert_hypervelocity_2017} for a discussion), however these preferentially eject younger stars and so observing stars which are now RRLs would only be possible if the Magellanic Clouds are bound to the Milky Way. It has been speculated that some OB stars observed to be coincident with the Magellanic Stream were born from the gas in the stream \citep{casetti-dinescu_constraints_2012}, however that cannot explain RRLs which have a typical age of around $10\;\mathrm{Gyr}$.

We simulate the interaction of the LMC and SMC during their first passage of the Milky Way in a N-body model based on the simulations used in \citet{belokurov_clouds_2017}. As in that work, we generate initial conditions for the LMC and SMC by sampling from the Clouds' observed proper motions \citep{kallivayalil_third-epoch_2013}, radial velocities \citep{vandermarel_2002,harris_zaritsky_2006}, and distances \citep{lmc_dist,smc_dist}. For each set of initial conditions, we evolve the LMC and SMC in the presence of a 3 component model for the Milky Way given by the \texttt{MWPotential2014} in \citet{galpy}. The LMC is modeled as a Hernquist profile \citep{hernquist} with a mass of $2.5\times10^{11} M_\odot$ and a scale radius of 25 kpc. The SMC is modelled as a Plummer sphere \citep{plummer} with a mass of $2\times 10^8 M_\odot$ and a scale radius of 1 kpc. The SMC's disruption around the LMC is simulated using the Lagrange point stripping technique described in \citet{gibbons_stream}. Given each set of initial conditions, the LMC and SMC are rewound from their current positions for 3 Gyr and then the disruption of the SMC is followed until the present. We ran 10000 such simulations with no further constraints on the debris properties. We also integrated a random sample of these orbits forward $1\;\mathrm{Gyr}$ into the future, and these orbits are shown in several plots throughout this work.

\begin{figure}
	\includegraphics[width=0.5\textwidth,trim =0mm 0mm 10mm 10mm, clip]{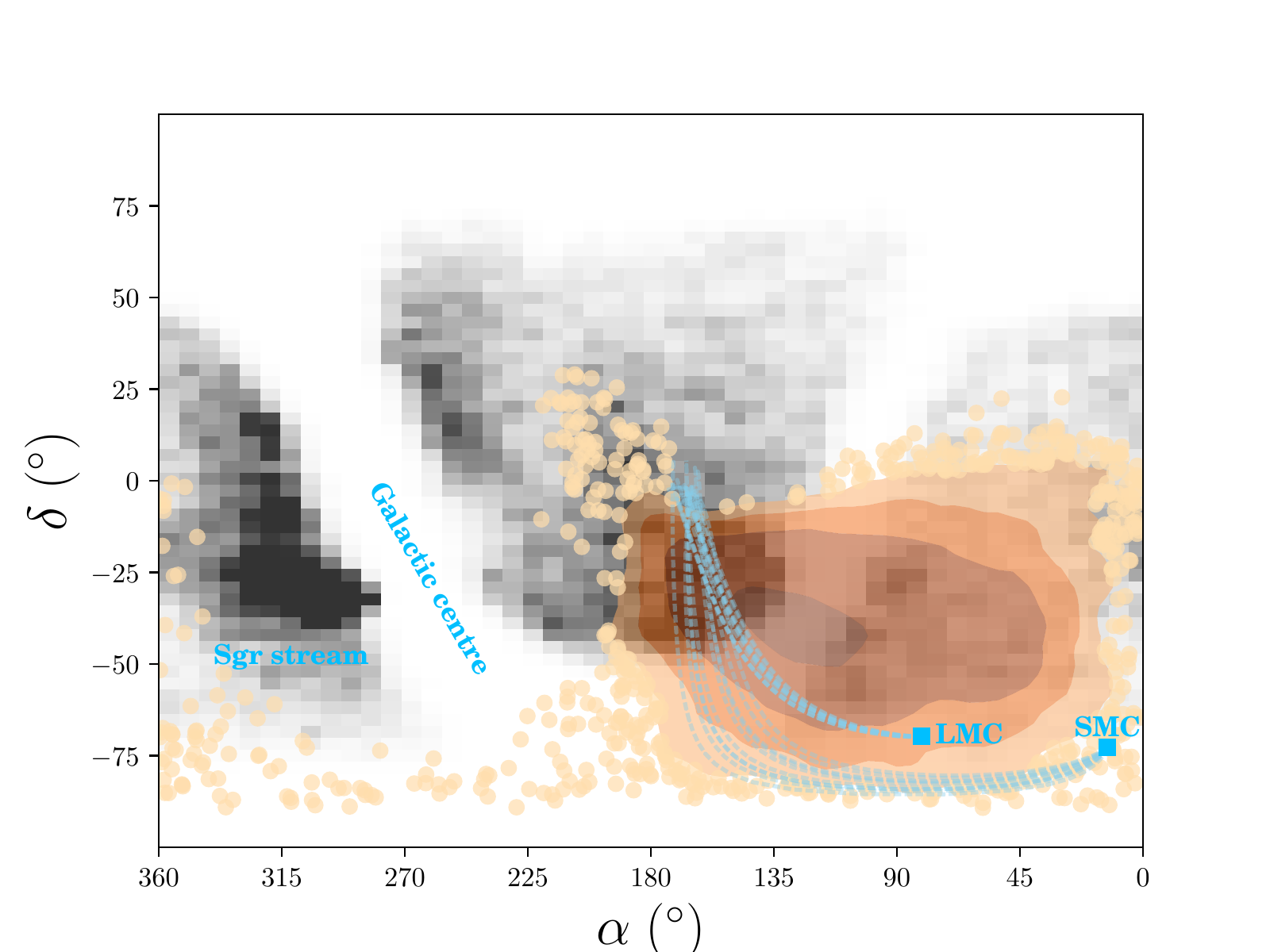}
	\caption{The underlying density and labels are the same as Fig. \ref{fig:crtsradec}. The contours and points are the locations of SMC debris particles from a simulation described in Section \ref{sec:magstream}.}
	\label{fig:crtssimradec}
 \end{figure}

\begin{figure}
	\includegraphics[width=0.5\textwidth,trim =0mm 0mm 15mm 12mm, clip]{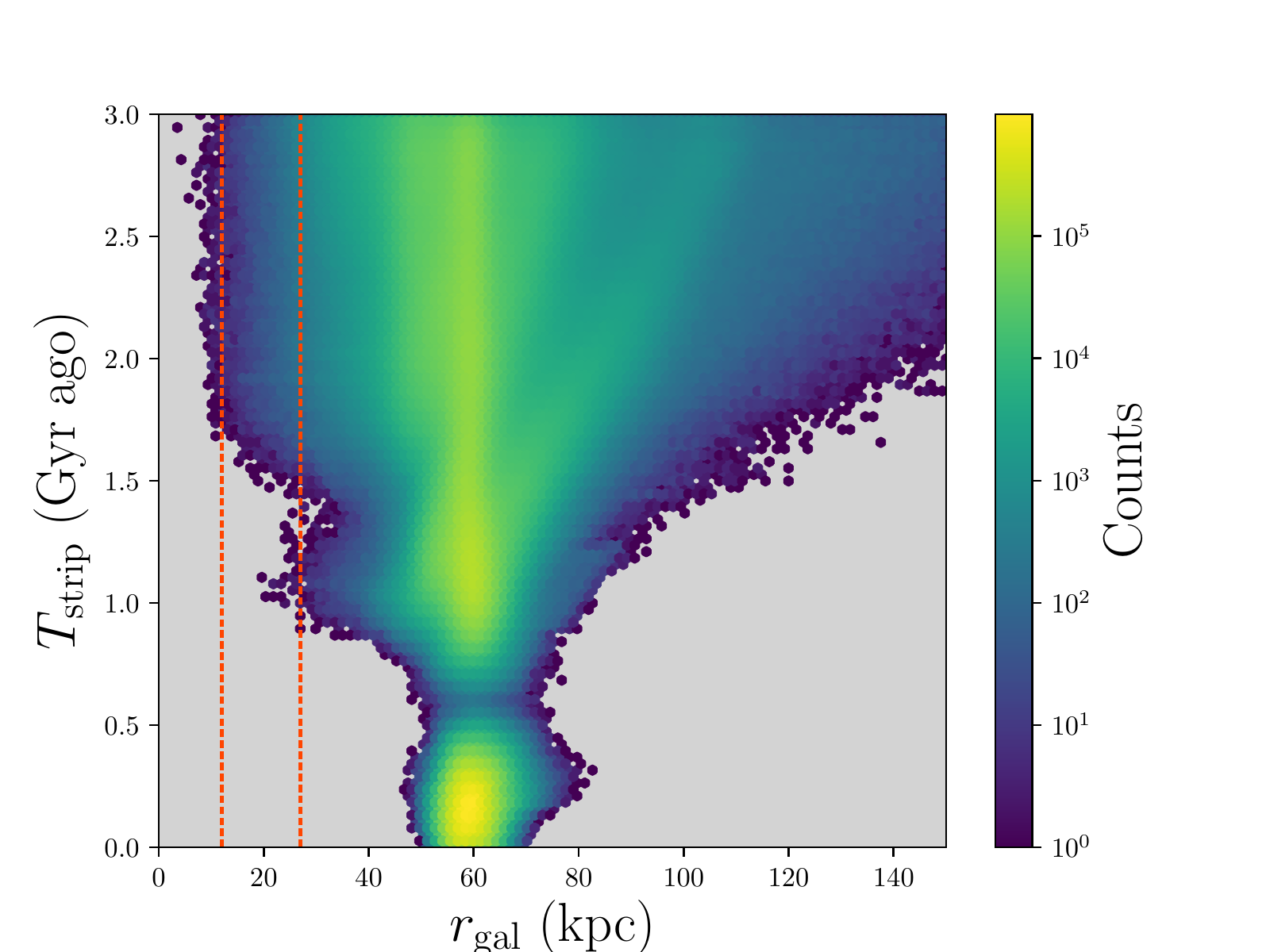}
	\caption{The time that simulated debris was stripped from the SMC versus present day Galactocentric radius.}
	\label{fig:rgaltstrip}
 \end{figure}

The results of these simulations are amalgamated and the particles representing SMC debris are shown in Figure \ref{fig:crtssimradec}, where we only show the particles that reach Galactocentric radii $12<R<27\;\mathrm{kpc}$. We find that SMC debris is stripped during either three or four pericentric passages of the LMC, but the debris that gets to a Galactocentric radius $12<R<27\;\mathrm{kpc}$ at present day has mostly been stripped more than $1.5\;\mathrm{Gyr}$ ago (see. Fig. \ref{fig:rgaltstrip}). This means that the present day distribution of nearby SMC debris is sensitive to the assumed orbital history of the Magellanic Clouds. For instance, Model 2 of \citet{besla_origin_2013} also had debris aligned with the structure, but the nearest of their debris lay in the range $30<R<40\;\mathrm{kpc}$. \citet{2010ApJ...723.1618N} estimated that the age of the Magellanic Stream is between $2.5\;\mathrm{and}\;3\;\mathrm{Gyr}$. If the RRLs and the Magellanic Stream left the Magellanic Clouds concurrently that would explain their similar extent.

\begin{figure}
	\includegraphics[width=0.5\textwidth,trim =0mm 0mm 0mm 0mm, clip]{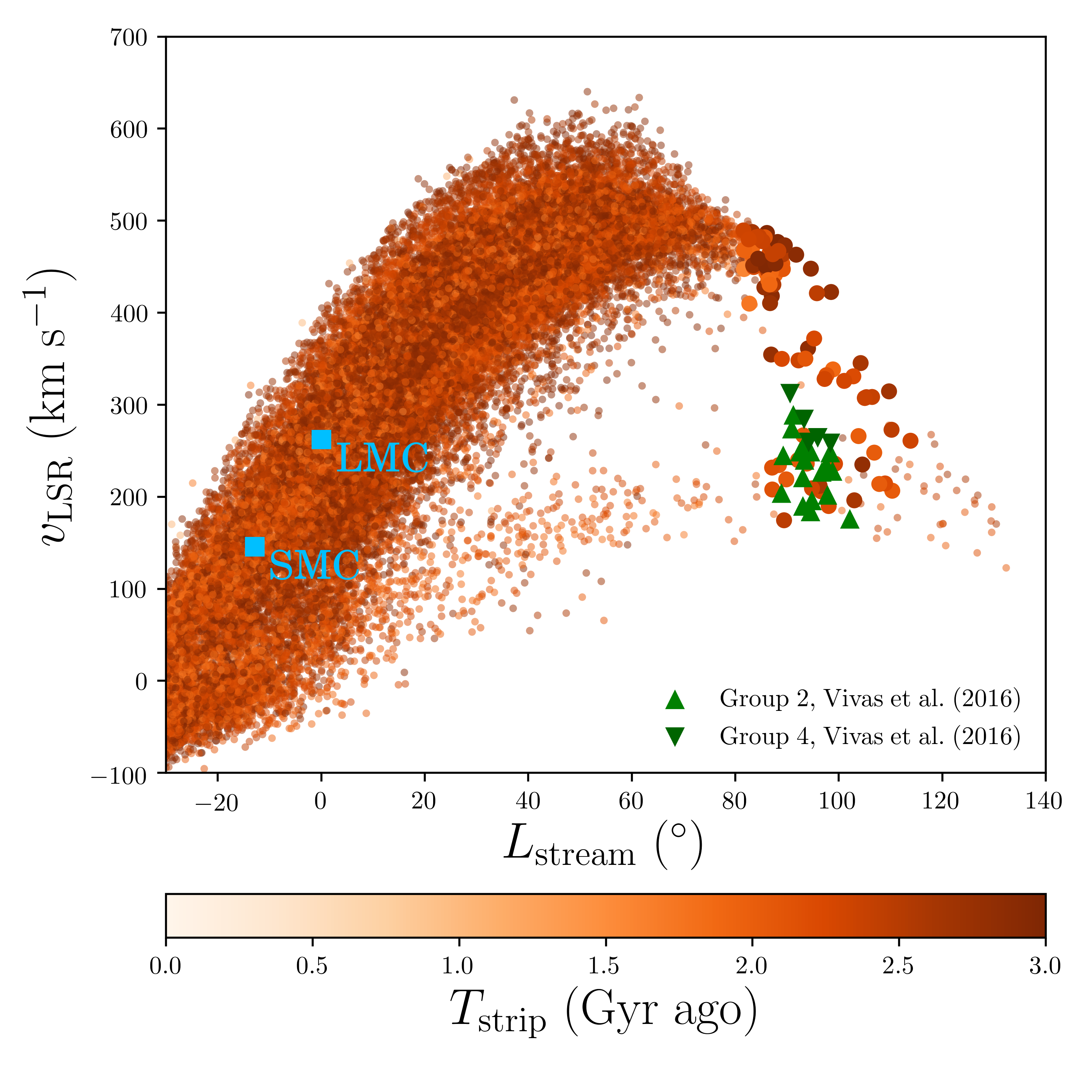}
	\caption{The longitude-velocity distribution of the simulated SMC debris with $12<R<27\;\mathrm{kpc}$. The colourscale gives the time $T_{\mathrm{strip}}$ that each particle was stripped from the SMC.  The debris which falls in the \citet{vivas_disentangling_2016} window is enlarged. Also shown are the two groups (2 \& 4) which \citet{vivas_disentangling_2016} argued were associated with the VSS.}
	\label{fig:vivas}
 \end{figure}

It is interesting to compare the simulated SMC debris to the \citet{vivas_disentangling_2016} RRL sample, because - while it is limited in angular coverage $175^{\circ}<\alpha<210^{\circ}$, $-10^{\circ}<\delta<10^{\circ}$ -  it should cut across a stream of stars from the Magellanic Clouds if it exists and all of their stars have radial velocities and distances. Note that the completeness of the \citet{vivas_disentangling_2016} sample is spatially heterogeneous and so we limit our comparison to the groups they explicitly associate with the VSS. In Figure \ref{fig:vivas} we show Groups 2 and 4 from \citet{vivas_disentangling_2016} together with our simulated SMC debris. The debris splits into two tracks: a primary one which contains most of the debris and spans all stripping times, and a secondary one which only features debris with $2.0<T_{\mathrm{strip}}<2.4\;\mathrm{Gyr}$. The secondary track of simulated SMC debris intersects with the location of these two groups which suggests that the VSS may simply be one component of a much larger complex of SMC debris. There are \citet{vivas_disentangling_2016} stars elsewhere in this space which may also be related to this structure but were not included in Groups 2 and 4 due to the strict membership criteria.

Only 4539 of our simulations produce stars in the radial range $12<R<27\;\mathrm{kpc}$ and only 37 of these produce debris in the lower of the tracks shown in Fig. \ref{fig:vivas}. This raises the possibility of using the sub-structure in Virgo to constrain the interaction history of the Magellanic Clouds. Each of our simulations corresponds to a random sample from the PDFs of the LMC's and SMC's distance, radial velocity and proper-motions and we investigated whether the initial conditions corresponding to these 37 simulations were clustered, however there was no clear evidence of this. One explanation is that whether SMC tidal debris reaches the inner Galaxy is determined by the LMC-SMC interaction history, and that history is highly sensitive to the initial conditions in a chaotic way. A detailed investigation should reveal constraints on the interaction history, but such an investigation is beyond the scope of this work.

\section{Conclusions}
\label{sec:conclusions}
We have mapped out a halo substructure in the RRL \emph{Gaia}+2MASS sample of \citet{iorio_first_2017}, the CRTS and the \citet{2016ApJ...817...73H} sample of RRLs in Pan-STARRS. These three samples have distinct systematics and so the existence of this halo substructure is secure.

The structure covers $\pi/4$ of the sky, occupies the distance range $12<R_{\mathrm{GC}}<27\;\mathrm{kpc}$, and is in near alignment with the Magellanic Stream. This structure has been previously discussed in the literature as the Virgo Over-Density or Virgo Stellar Stream, and thus may be debris from a disrupted dwarf galaxy. 

An alternative, novel explanation is that these stars have been stripped from the halo of the SMC by the tidal action of the LMC during a close encounter about $3\;\mathrm{Gyr}$ ago. Simulations of the SMC-LMC interaction commonly have particles in this region of the sky, although the requirement that they reach $12<R_{\mathrm{GC}}<27\;\mathrm{kpc}$ places a strong constraint on the configuration of that interaction. This would imply the Virgo sub-structure is Magellanic in origin and we have shown that this structure can be replicated in simulations of SMC debris. \citet{2014A&A...566A.118D} measured the metallicity of the VSS to be $[\mathrm{Fe}/\mathrm{H}]=-1.78\pm0.37\;\mathrm{dex}$ which is in accord with the metallicity of the SMC, $[\mathrm{Fe}/\mathrm{H}]=-1.70\pm0.27\;\mathrm{dex}$, estimated from the lightcurves of 1831 RRLs \citep{2012AJ....143...48H}.

\citet{iorio_first_2017} conclude that by applying their method to the second data release (DR2) of \emph{Gaia} they will obtain an all-sky RRL sample that stretches out to $100\;\mathrm{kpc}$. Proper motions \hl{are} included in \emph{Gaia} DR2 for all these RRLs and these can be converted to tangential velocities using the distance estimates. The Magellanic Clouds are on a polar orbit and have large tangential velocities \citep{kallivayalil_third-epoch_2013}, thus the tangential velocities of RRLs in the structure will test whether they are associated with the Clouds and, if so, strongly constrain the Magellanic Clouds' interaction history. If they are not associated with the Magellanic Clouds then the tangential velocities will probe the orbit of the disrupted progenitor dwarf galaxy. \hl{Future exploration of} \emph{Gaia} will reveal the full intricacy of the Milky Way halo and may uncover a leading stellar arm of the Magellanic Clouds much nearer than previously thought possible.

While we were revising this work, others have begun the task of studying the substructure in Virgo through the lens of \textit{Gaia} DR2. \hl{\citet{2018arXiv180701335S} cross-matched RRLs associated with the Virgo (412 RRLs from \citealp{vivas_disentangling_2016}) and Hercules-Aquila (46 RRLs from \citealp{2018MNRAS.476.3913S}) sub-structures with \textit{Gaia} to obtain their proper motions, and then calculated their orbital properties in the Galactic potential. \citet{2018arXiv180701335S} show that the stars in Group 2 of \citet{vivas_disentangling_2016} are on highly eccentric orbits and have mean apocentric radii between $20$ and $40\;\mathrm{kpc}$, perhaps ruling out a Magellanic origin for these stars. However, the range of apocentric radii derived by \citet{2018arXiv180701335S} for the entire Virgo RRL sample extends out to $60\;\mathrm{kpc}$ and thus a Magellanic origin is plausible for at least some of these stars.\citet{2018arXiv180701335S} demonstrated that the distribution of eccentricities, pericentres and apocentres of the stars in the VOD that have Galactocentric radii $11<R<16\;\mathrm{kpc}$ (the range covered by the Hercules-Aquila Cloud) qualitatively matches the distributions of stars in the Hercules-Aquila Cloud; the authors argued that this is evidence for a common origin in the ancient massive merger of the recently identified `Sausage' galaxy \citep{2018MNRAS.478..611B}. If all the sub-structure towards Virgo does originate in this merger, then that would preclude our Magellanic hypothesis. However, as discussed earlier in this work, the sub-structure towards Virgo is complex and may contain multiple structures with distinct origins (e.g. the VOD is from the `Sausage' merger while the VSS is debris from the Magellanic Clouds).}

\section*{Acknowledgements}
The authors are grateful to the reviewer for in-depth comments that greatly improved the focus of this paper. DB is grateful to the Science and Technology Facilities Council (STFC) for providing PhD funding. The research leading to these results has received funding from the European Research Council under the European Union's Seventh Framework Programme (FP/2007-2013) / ERC Grant Agreement n. 308024. This work has made use of data from the European Space Agency (ESA) mission {\it Gaia} (\url{https://www.cosmos.esa.int/gaia}), processed by the {\it Gaia} Data Processing and Analysis Consortium (DPAC, \url{https://www.cosmos.esa.int/web/gaia/dpac/consortium}). Funding for the DPAC has been provided by national institutions, in particular the institutions participating in the {\it Gaia} Multilateral Agreement. This publication makes use of data products from the Two Micron All Sky Survey, which is a joint project of the University of Massachusetts and the Infrared Processing and Analysis Center/California Institute of Technology, funded by the National Aeronautics and Space Administration and the National Science Foundation. The CSS survey is funded by the National Aeronautics and Space Administration under Grant No. NNG05GF22G issued through the Science Mission Directorate Near-Earth Objects Observations Program.  The CRTS survey is supported by the U.S.~National Science Foundation under grants AST-0909182 and AST-1313422. Funding for SDSS-III has been provided by the Alfred P. Sloan Foundation, the Participating Institutions, the National Science Foundation, and the U.S. Department of Energy Office of Science. The SDSS-III web site is \url{http://www.sdss3.org/}. SDSS-III is managed by the Astrophysical Research Consortium for the Participating Institutions of the SDSS-III Collaboration including the University of Arizona, the Brazilian Participation Group, Brookhaven National Laboratory, Carnegie Mellon University, University of Florida, the French Participation Group, the German Participation Group, Harvard University, the Instituto de Astrofisica de Canarias, the Michigan State/Notre Dame/JINA Participation Group, Johns Hopkins University, Lawrence Berkeley National Laboratory, Max Planck Institute for Astrophysics, Max Planck Institute for Extraterrestrial Physics, New Mexico State University, New York University, Ohio State University, Pennsylvania State University, University of Portsmouth, Princeton University, the Spanish Participation Group, University of Tokyo, University of Utah, Vanderbilt University, University of Virginia, University of Washington, and Yale University. The Pan-STARRS1 Surveys (PS1) and the PS1 public science archive have been made possible through contributions by the Institute for Astronomy, the University of Hawaii, the Pan-STARRS Project Office, the Max-Planck Society and its participating institutes, the Max Planck Institute for Astronomy, Heidelberg and the Max Planck Institute for Extraterrestrial Physics, Garching, The Johns Hopkins University, Durham University, the University of Edinburgh, the Queen's University Belfast, the Harvard-Smithsonian Center for Astrophysics, the Las Cumbres Observatory Global Telescope Network Incorporated, the National Central University of Taiwan, the Space Telescope Science Institute, the National Aeronautics and Space Administration under Grant No. NNX08AR22G issued through the Planetary Science Division of the NASA Science Mission Directorate, the National Science Foundation Grant No. AST-1238877, the University of Maryland, Eotvos Lorand University (ELTE), the Los Alamos National Laboratory, and the Gordon and Betty Moore Foundation.

%%%%%%%%%%%%%%%%%%%%%%%%%%%%%%%%%%%%%%%%%%%%%%%%%%



\bibliographystyle{mnras}
\bibliography{reference} % if your bibtex file is called example.bib

%\begin{thebibliography}{99}
%\end{thebibliography}

%%%%%%%%%%%%%%%%%%%%%%%%%%%%%%%%%%%%%%%%%%%%%%%%%%

%%%%%%%%%%%%%%%%% APPENDICES %%%%%%%%%%%%%%%%%%%%%

%\appendix

%%%%%%%%%%%%%%%%%%%%%%%%%%%%%%%%%%%%%%%%%%%%%%%%%%

% Don't change these lines
\bsp	% typesetting comment
\label{lastpage}
\end{document}